# Back-Scattering Suppression for Broad-Spectral High-Absorption Silicon Extended Area Blackbody

Hong-Shuai Zhou[1,†], Jin-Hao Zhang[2,†], Ben-Feng Bai[1,*], Xi-Ran Mei[1], Kun-Peng Chen[1], Xiao-Peng Hao[1,3,4,*], Jian Song[3,4], Guo-Rui Guo[3,4], Jia-Lin Chen[2], Tian Tian[2], Wan-Jie Shen[2], Zi-Heng Zhong[2], Jia-Yao Liu[2], Ji-Hong Zhao[5] and Hong-Bo Sun[1,5,*]

[1]State Key Laboratory of Precision Measurement Technology and Instruments, Department of Precision Instrument, Tsinghua university, Beijing 100084, China.
[2]Shenzhen International Graduate School, Tsinghua university, Shenzhen 518055, China.
[3]Remote Sensing Calibration Laboratory, National Institute of Metrology, Beijing 100029, China.
[4]Technology Innovation Center of Infrared Remote Sensing Metrology Technology, State Administration for Market Regulation, Beijing 100029, China.
[5]State Key Laboratory of Integrated Optoelectronics, College of Electronic Science and Engineering, Jilin University, Changchun 130012, China.

[†]These authors make the same contribution to this work.

[*]Corresponding author e-mail:
　　baibenfeng@tsinghua.edu.cn
　　haoxp@nim.ac.cn
　　hbsun@tsinghua.edu.cn

## 0. Abstract

The stability and emissivity of the online calibration blackbody used in high-precision infrared remote sensing detectors in extreme environments are the primary limiting factors for their measurement accuracy. Due to the limitations of microstructure size effects, traditional calibration extended area blackbody cannot achieve an optimal balance between emissivity and stability, thus hindering further improvement in infrared remote sensing accuracy. This work proposes a new method that utilize suppressing near-field backscattering to control far-field reflectance. Specifically, through simultaneously reducing backscattering intensity and the backscattering solid angle, the reflectance is significantly reduced to an extremely low limit, which is validated through numerical simulations. Additionally, by combining the femtosecond laser self-convergent processing technique, the spontaneous energy negative feedback mechanism during femtosecond laser processing is utilized to achieve the fabrication of a high emissivity, thermally stable, mechanically stable, and highly uniform extended area blackbody. The blackbody fabricated using this technique can be applied for online calibration in various extreme environments, significantly improving measurement accuracy and service life.

## 1. Introduction

Remote sensing and measurement technologies have found widespread applications in various fields, such as earth observation[1-3], meteorological forecasting[4-6] and satellite navigation[7-9], owing to theirs non-contact characteristics. One of the most common methods of implementing remote sensing is through the detection of electromagnetic waves[10-13], which enables measurement of the target object or environment. According to Planck's law, any object above absolute zero emits infrared radiation, and its emission spectrum and energy distribution are related to the object's temperature[14-18]. This characteristic of thermal radiation makes space infrared remote sensing possible[19-22].

Infrared remote sensing offers several advantages over other methods, primarily due to the temperature dependence of radiation. In addition to imaging capabilities, infrared remote sensing allows for the measurement of temperature at any point within the imaging field through the inversion of radiative energy[23-25]. This makes it more advantageous for precise temperature measurements. However, for infrared remote sensing to be effective, not only the system is capable of detecting objects at a long distance, but it is the accuracy of detection system that must be ensured. The key to achieving this lies in the calibration and validation of both the radiative properties and the detector responses. Typically, calibration is performed using extended area blackbody with known temperature and radiative properties, providing online imaging calibration to achieve high-precision measurements.

Due to the extreme and harsh conditions of the operating environment, high-performance online calibration blackbodies must have excellent thermal stability, great mechanical stability, high emissivity, and large surface areas with minimal spatial non-uniformity. According to Kirchhoff's law of radiation, the emissivity of an object's surface at a specific wavelength equals its absorptivity at that wavelength. Therefore, a key technical challenge is developing infrared high-absorption materials with excellent thermal stability, mechanical stability, and uniformity. Overcoming this challenge will directly enhance the accuracy of space remote sensing measurements in harsh environments, which has been the focus of considerable research.

Currently, two main approaches are commonly used to achieve high absorption: coatings and microstructural methods. The coating method involves applying materials with high emissivity to the surface of a substrate. However, this approach is highly influenced by the material properties, leading to challenges in achieving long-term stability. On the other hand, microstructural methods have been extensively researched in recent years to achieve high absorption[26-39]. This approach involves designing and fabricating microstructures using micro-nano fabrication techniques to achieve broadband absorption properties. Typical examples include carbon nanotubes and black silicon. These materials overcome impedance mismatch and exhibit good infrared absorption properties due to their dense nanostructures. However, their performance is compromised by the high-temperature instability and relatively poor mechanical stability caused by the nanoscale structures.

The use of femtosecond laser processing to fabricate micro-nano composite structures on substrates has been proposed as a potential solution[40-51]. This method can produce structures with both high emissivity and improved stability, offering a promising solution to the challenges mentioned above. However, current theories regarding optical absorption in micro-nano composite structures primarily rely on qualitative analysis of geometric light trapping[44, 48, 51]. As the size of the microstructures approaches the wavelength of light, the geometric light trapping effect becomes ineffective, leading to a lack of theoretical understanding and design guidelines for achieving better emissivity and stability, therefore, it is necessary to propose new theories beyond geometric optics. Moreover, the limitations of current theories make it difficult to fabricate blackbodies with better comprehensive performance, so it is urgent to develop a new technique for fabricating extended area blackbodies with excellent performance based on the new theories.

To address these issue, this work proposes a far-field reflectance control model based on particle backscatter suppression. A femtosecond laser processing technique with self-convergence is used to fabricate the designed structures on silicon surfaces. The fabricated structures not only exhibit very high emissivity but also demonstrate excellent thermal stability, mechanical stability, and uniformity. Notably, these structures maintain emissivity uniformity even after exposure to extreme environments at 900 °C. Moreover, the technique allows for the fabrication of large area structures, effectively resolving the calibration issues encountered in infrared remote sensing measurements in extreme environments.

## 2. Materials and Method

### 2.1. experimental setup

In this work, we use 700 μm, <100> single-side polished N-type doped silicon as the processing substrate, which has no transmission in the range of 3~14 μm; In the thin-layer absorption test, we use 500 μm, <100> double-sided polished intrinsic silicon as the processing and test substrate, which has a high transmission rate of 3~14 μm. All silicon wafers are cut into 30 mm squares. Fiber femtosecond laser (FemtoYL-20) from YSL Photonics is used for processing. The repetition rate is 25 kHz, the maximum single pulse energy is 200 μJ, the central wavelength is 1030 nm, and the pulse width is 300fs. The expanded beam has a diameter of 5 mm (light intensity reduces to $e^{-2}$ of peak position ) and is focused on the sample surface through a Mitoyo flat-field apochromatic objective with a numerical aperture of 0.14. We use confocal systems and industrial cameras to image the processing in real time, and ensure that the laser used for processing is linearly polarized with the polarization direction parallel to the X-axis of motion before processing, as shown in Fig. S2.

### 2.2. Numerical simulation

We used the wave optical module of COMSOL Multphysics combined with the far-field domain to numerically simulate the far-field scattering and reflection characteristics of particles and thin films. The boundary conditions were Perfectly Matched Layer (PML) and an Perfect Electrical Conductor (PEC), and the simulation region was set to the far-field domain. The results were shown in Fig. 1a~e. The solid-state heat transfer module of COMSOL Multphysics was used to conduct thermodynamic simulation on the cross section of laser influence region. The set single pulse energy was 10 μJ, beam diameter was 10 μm, action time was 300fs, and surface laser reflectance was 0.3. The results were shown in Fig. 2i~j and Fig. 3a~e.

### 2.3. Sample fabrication

The three-axis displacement stage used is from ACS Motion Control with an accuracy of 100 nm. Before processing, the samples were ultrasonic cleaned with acetone, ethanol and deionized water sequentially. During processing, the three-axis displacement stage drives the sample to move at a certain speed and direction, and produces relative displacement with the laser. The surface processing path was a grid structure, the grid period was 30 μm, and the processing speed was 50 mm/s (Fig. 2 sample) and 10 mm/s (Fig. 1 sample). All sample processing areas are 23 mm in size to meet the characterization requirements of all characterization instruments. In the high temperature test, the samples were treated at high temperature by using a heated high temperature muffle furnace. The temperature rise took 5 h, followed by a 5-hour holding time at the target temperature, and then cooled to room temperature naturally. In the stripping experiment, the samples were fully covered and pasted with 3M tape, then forced to peel, then soaked in anhydrous ethanol for 1 minute, and finally washed with deionized water.

## 2.4. Sample characterization

The low-temperature vacuum Fourier transform infrared spectrometer (FTIR, BRUKER GMBH, VERTEX 70v) and integrating sphere were used to measure the reflectance of the samples. The test equipment and work were provided by National Institute of Metrology. The surface morphology and elemental composition of the samples were measured by cold field emission scanning electron microscope (FE-SEM, JEOL, JSM-IT700) and energy dispersive spectrometer (EDS, JEOL, JSM-IT700). X-ray photoelectron spectroscopy (XPS, Thermo Fisher Scientific, 250XI)was used to measure the exact composition and chemical state of the elements on the surface of the samples, which has an energy resolution of less than 0.45 eV. The deep level transient spectrometer (DLTS, Phys Tech GmbH, HERA DLTS FT-1030) was used to measure the deep level near the surface. The homogeneity of the samples was measured by an infrared thermal imager (DALI TECHNOLOGY, DM60) operating in 8~14 μm.

## 3. Result and discussion

## 3.1. The principle of near-field backscatter determines the far-field reflectivity

When a beam of light passes through an interface made of two materials, its reflectance is determined by Fresnel's law. Fig. S1 shows a Fresnel reflectance simulation when materials with complex refractive index are put in air and a beam of light passes through it. We can see that the interface reflectance is higher because of Fresnel's law and light impedance mismatch. It needs the real part of index less than 1.5 and the imaginary part of index less than 0.4 to get low reflectance less than 5%, and the lower reflectance require the less index. Thus for material with specific application fields, the interface optical reflection is astricted by

Fresnel's law and cannot be reduced.

To bypass the astrict of Fresnel's law, we consider a particle scattering system. Fig. 1a is the schematic diagram, Fig. 1b is the simulation diagram. When a beam of light irradiates to a particle, the scattering occurs, composed of forescatter and backscatter. The simulation diagram shown in Fig. 1b calculates the integral scattering cross sections of different particle sizes in 2π solid Angle in the opposite direction of light propagation, which are equal to the total backscatter rates. Fig. 1a shows the differential direction scattering cross sections of small (0.2 μm), medium (0.5 μm) and large (1.2 μm) silica particles sizes in parallel and perpendicular light polarization directions under 5 μm infrared light irradiation (the half part of left and right). It shows that the backscatter cross section is strongly logarithmically correlated with the size of particles, and small particles have a smaller differential scattering cross section. If the whole surface is randomly tiled with controllable particle size, the near-field backscatter determines the far-field reflectivity, so the interface reflectance can be flexibly adjusted by controlling the tiled particle size.

Furthermore, Fig. 1c~d give the simulation results of integrated backscattering cross sections of N-type doped silicon and silicon dioxide spheres at 3~14 μm by using the method as shown in Fig. 1b. The solid line represents the 5% backscattering rate boundary, and the dashed line represents the 1% boundary. The trend of integral simulation results is the same as that of differential simulation results. For both silicon and silica, the backscattering can be reduced to 1% when the particle size is around 200 nm, indicating that the method has a certain non-material sensitivity to reflection regulation. For comparison, Fig. 1e gives the variation of reflectivity of silicon dioxide film on silicon with different dioxide thickness. It is obviously that the reflectance of thin films is much higher than that of particles at the same scale. The above theoretical calculation results show that determining the far-field reflectivity by near-field backscatter control can significantly suppress the Fresnel reflection of the interface.

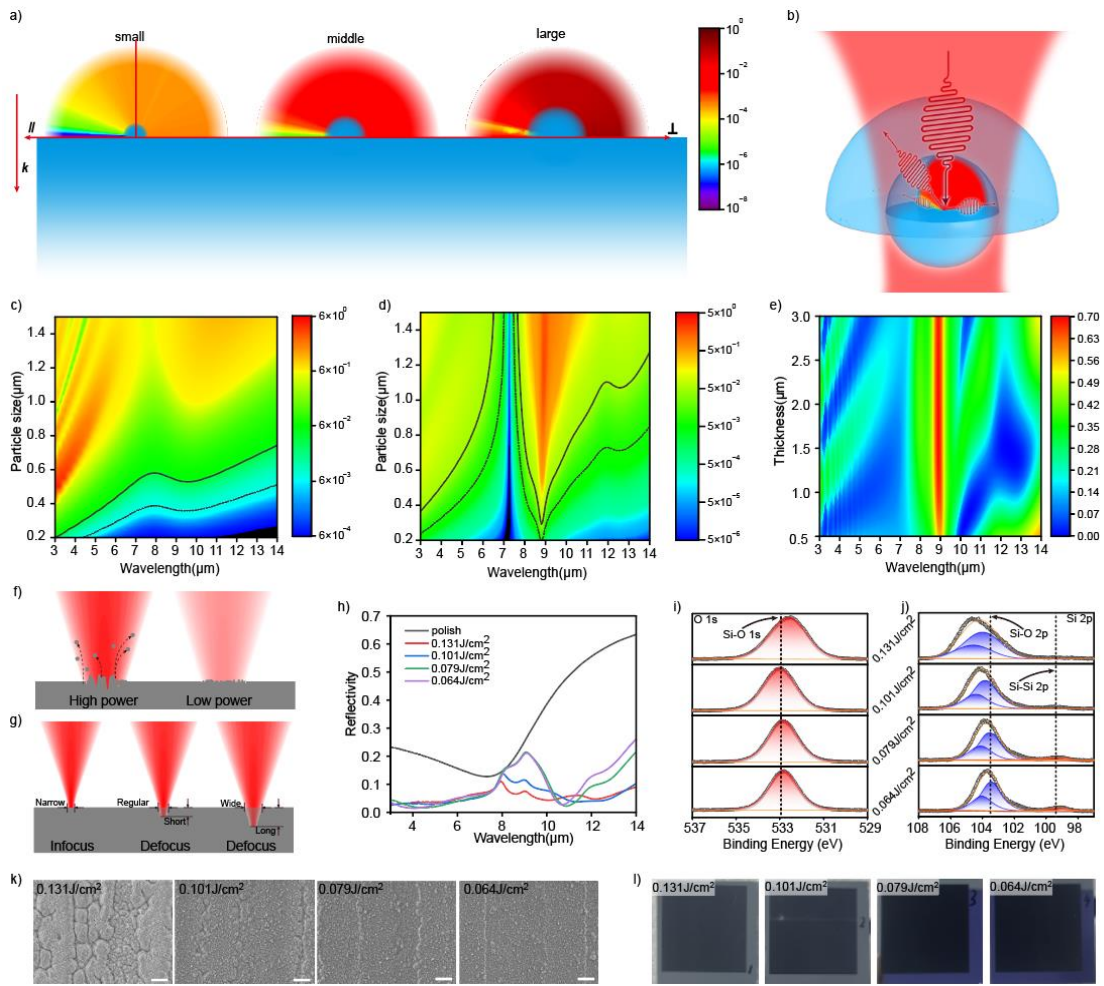

Fig1 a) Diagram of near field scattering suppress far field reflection. The colorful regions above the particles are the differential scattering cross section of different backscattering angles, and the left and right half-sides of the particle represent the differential scattering cross section in the quarter plane parallel to and perpendicular to the laser polarization. b) Simulation diagram. The large hemisphere represents the scattering collection area, and the small sphere is the scatterer. c ~ d) The integral backscattering cross sections of N-Si and $SiO_2$ vary with particle size and wavelength. The solid and dashed lines indicate the 5% and 1% reflectance boundaries. e) The reflectance of $SiO_2$ film on silicon varies with film thickness and wavelength. f ~ g) Principle of laser process. f) Control the energy density to generate low-size particles. g) Defocus to control power density. h) Reflectance results at different energy densities. i ~ j) XPS data of the surface of samples processed with different energy densities. k ~ l) SEM images (10 μm

**scale) and physical photos of the surface of samples processed with different energy densities.**

**3.2. Ultrafast laser control particle size**

Although the above analysis and results show that a random distribution of particles of a specific size on the surface can significantly reduce reflection, generating random particles with strong adhesion of a specific size is a hard challenge. Although lithography and nanoimprint, as the most commonly used micro-nano processing methods, can generate such size of structures, they cannot achieve true random distribution and spherical particles spreading over the entire surface. Dry etching processes such as Focused Ion Beam (FIB) etching and Inductively Coupled Plasma (ICP) etching cannot generate controllable surface particles and are strongly dependent on templates. Wet etching process can generate lots of nanostructures in a short time, but it is strongly dependent on the material lattice and cannot form approximately spherical particles. Physical vapor deposition (PVD) technology can generate uniform spherical particles on the surface, but it is difficult to control their size.

Because normal micro and nano processing methods cannot achieve controllable distribution of random particle size, we use ultrafast laser surface processing technology. Ultrafast laser surface processing is a new treatment technology emerging in recent years. The schematic diagram of its processing system is shown in Fig. S2. Picosecond or femtosecond laser is finally focused by the objective lens to the surface of the sample after passing through the mirror group, and the sample is moved by a high-precision displacement stage to achieve surface processing on a specific path.

As shown in Fig. 1f, the energy density of femtosecond laser at the focus position is higher, and the violent interaction with the material usually results in obvious material removal and chipping, which is impossible to achieve controllable particle distribution. Therefore, accurate regulation of femtosecond laser energy density is the key to generate particle size distribution. Here, we change the beam size by precisely controlling the vertical displacement between the focus and the processed surface, thus transforming the precise regulation of the laser energy density into the precise regulation of the focus position, and then to achieve controllable particle size distribution. The schematic diagram of the principle is shown in Fig. 1g.

The detailed control principle and calculation method for energy intensity are shown in Fig. S3. The sample used is N-type silicon; Fig. 1l shows the surface of the sample treated by femtosecond laser with four different energy densities; Figure 1k is the SEM image(scale 10 μm) of the corresponding sample; Figure 1h is the reflectance result of the corresponding sample(the method of reflectance testing is described in S4),in which the polish is an unprocessed silicon sample with a smooth surface. Fig1. i ~ j shows the XPS test result of the sample. It shows that the surface of the processed sample is composed of Si and $SiO_2$ particles (mostly $SiO_2$), so the surface optical properties are regulated by both Si and $SiO_2$ particles. As is seen from Fig. 1d, $SiO_2$ mainly affects the reflectance of 8~10 μm. However, the reflectance trend of Si is relatively flat in the whole band range of 3~14 μm, which is verified by the data in Figure 1h, and the relatively high reflectance in 8~10 μm is caused by $SiO_2$. It can be seen that by reasonably controlling the laser processing parameters, small particles with uniform distribution can be generated on the flat surface within a certain range of laser energy density, and the size is less than 500 nm. By this method, the sample with much smaller reflectance than the polished plane in a broadband can be fabricated, and its reflection intensity basically meets the prediction in Fig. 1c~d. By using defocus method, the processing has great robustness, and the results shown in Fig. 1k~l are processed at 700, 800, 900, 1000 μm defocus respectively (corresponding energy densities are 0. 131,0.101,0.079,0.064, and Fig. S3 gives the calculation method).

### 3.3. Microstructure reduces the integral Angle of the backscattering cross section of the particles

Although the reflection can be reduced by controlling the particle size, it is extremely difficult to further reduce the reflectivity due to the difficulty of controlling the particle size and distribution very precisely. The total backscattering intensity of particles is determined by the following formula:

$$I = \iint_\Omega I(\theta, \varphi) \mathrm{d}\theta \mathrm{d}\varphi \qquad (E1)$$

Where $I(\theta, \varphi)$ is the intensity of the scattered light in the direction of $\theta$ (azimuthal angle) and $\varphi$ (zenith angle) and $\Omega$ is the integral solid angle of the backscattering.

Reducing the far-field reflection by decreasing the particle size is essentially reducing the backscattering intensity per unit solid angle $I(\theta, \varphi)$. We consider another factor that affects the overall strength of backscattering: the integral solid angle of backscattering $\Omega$. The integral solid angle of backscattering in the free space plane is $2\pi$. If we can reduce the integral angle of backscattering on the free space plane, the far-field reflection caused by backscattering can be further reduced. Therefore, we consider a physical model in which the sphere particles are randomly distributed on an inclined plane, as shown in Fig. 2a. Due to the shielding effect of the backscattering of particles by the inclined plane, the final integral angle of backscattering can be less than $2\pi$, and the far-field reflectance can be further reduced.

Based on this principle, we developed a nanoparticle backscattering suppression structure based on micron conical tips. By fabricating conical tips with larger structures and generating nanoparticles as shown in Fig. 1 on their surface, double suppression of backscattering can be achieved. The physical photo of the sample with micron cones and nanoparticles is shown in Fig. 2b, and the SEM image is shown in Fig. 2c (the inset shows its top view).

Fig. 2d~e shows the simulation results of Si and $SiO_2$ particles respectively under different inclined angles according to Formula E1. The inset in Fig. 2d shows a simulation diagram where the scattering within the colorful symmetric line segment contributes to the far-field reflection, while the light outside the line segment doesn't contribute to the far-field reflection due to the shielding effect of the inclined plane. The angle marked in the figure is the angle between two line segments ranging from 0~180°, and the simulated solid angle region is the corresponding three-dimensional rotating cone (Because the cone shown in Fig. 2c has rotational symmetry, it is reasonable to adopt a rotating cone in the back

integral region as the solid angle of the nanoparticles backscattering).

The simulation results show that the structure can inhibit the light reflection. The introduction of the scattering suppression microstructure can significantly reduce the surface reflectance compared with the plane, especially for the short-wave segment (3~6 μm) and the long-wave (8-10 μm) affected by $SiO_2$, when the top angle is 60° (corresponding to the base angle of the cone is 60° as well), it can reduce the reflectance to 1/3 of the plane structure (as shown in the green and black lines).

In order to obtain the structure of Fig. 2c, we still adopt the defocusing processing method shown in Fig. 1g. Different from the above processing parameters, we use a smaller defocusing amount (~150 μm, seen in S3) in this step to obtain a larger energy density and achieve the high-power processing effect shown in Fig. 1f. The detailed processing procedure is shown in Fig. 2f, where we propose a laser processing method to achieve structure stability and convergence through multiple processing times. When the ultrafast laser firstly acts on the surface of the sample, due to the violent interaction between the laser center with high energy density and the material, the material removal and large debris will be obvious, and the surface will form a shallow and rough structure. With the increase of the times of laser processing, the area of high laser energy will result in the corresponding material removal each time, which will lead to the improvement of the functional structure depth, but the material removal caused by the strong interaction is obvious, and the particle structure covering the surface is still large (5 ~ 10 times); With the further increase of processing times (10 ~ 20times), the material within the high energy density area of the laser center has been removed almost entirely, and the removal effect caused by the strong laser and material interaction disappears, but the weak energy density of the laser edge interacts with the edge of the removal structure, achieving the low-power processing effect as shown in Fig. 1f. Finally, smaller particles are generated on the surface of the structure. According to this principle, as long as there are enough processing times, the microstructure will always converge to our target structure. Fig. 2g shows the SEM images of the structure morphology corresponding to Fig. 2f, and its morphological changes are basically consistent with the above mechanism model. Through comparison of SEM images, it is found that the convergent processing times are about 15 times in the Si material system, and there will be no obvious morphological changes when further increasing the processing times, which further verifies the convergence of the structure with respect to the processing times.

For ordinary materials, due to Kirchhoff's radiation law, the emissivity of an object in thermal equilibrium is equal to its absorption rate, so the emissivity of our sample is calculated as follows:

$$\varepsilon = A = 1 - R - T \qquad (E2)$$

Where $\varepsilon$ is the surface emissivity, A is the optical absorption of the material, R and T are the reflectance and transmittance of the material surface. The silicon used is N-type doped silicon, which is opaque in the infrared band, i.e., T=0.

Fig. 2h shows the emissivity measurement curve of the processed surface morphology shown in Fig. 2g. With the increase of processing times, the emissivity of the surface gradually increases (that is, the reflectance gradually decreases), which is accordance with the variation law of reflectance predicted by Formula 1. Meanwhile, the 15 times (green) and 20 times (purple) curves in the emissivity data graph almost overlap in concern band. **The optical property convergence of the structure after 15 processing times is further verified.**

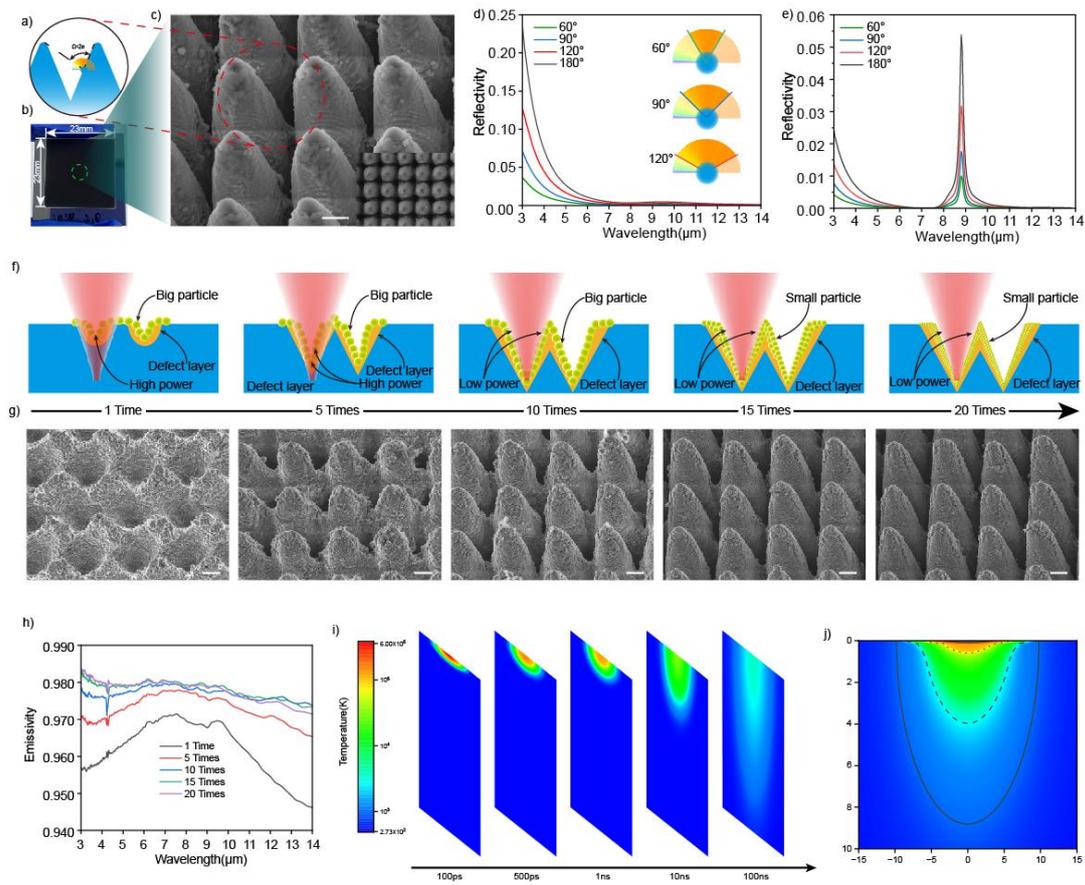

**Figure2 a)** The principle of microstructure reduces the integral solid angle. **b)** The sample photo and **c)** SEM enlarged view, the inset is the top view. **d)** and **e)** are the different microstructure cone angles reflectance simulation of Si and $SiO_2$, the inset of d) is the diagram that different cone angles reduce reflectance. **f)~g)** The schematic diagram of morphology change and convergence with processing times increasing and the SEM image. **h)** The measurement results of emissivity variation with wavelength and processing times. **i)** Thermal simulation. The change of temperature distribution with time in the cross section of laser interaction region. **j)** The maximum temperature at each point on the cross-section of the laser interaction region. The solid line, dashed line, dotted line are boundaries of 1000K, 10000K, 50000K respectively.

## 3.4. The formation of defect absorbing working medium under strong laser pulse

Although the above models and results can significantly reduce far-field reflection and contribute to achieving high absorption, they do not explain the physical mechanism of optical absorption. However, high absorptivity is a necessary and sufficient condition to achieve high emissivity. To elucidate the mechanism of optical absorption, we consider a simplified thermal process of a single femtosecond laser pulse interacting with the material. Fig. 2i shows the simulation diagram of temperature change inside the material with time under the influence of a Gaussian pulse at 10 uJ@300 fs. The waist radius of the focused Gaussian beam used in the simulation is 5 μm. Fig. 2j shows the maximum temperature at each point on the cross-section of the laser interaction region. The simulation results show that the maximum temperature of non-equilibrium state in the laser-affected cross-section can be close to $10^5$K. If taking 1000K as the laser thermal affected boundary, its depth can reach 10μm, indicating that high pulse energy has a significant thermal effect on the material.

The high temperature of the transient non-thermal equilibrium state does not cause significant melting and lattice destruction of the material, but changes the energy distribution of the atoms. As shown in Fig. 3a, the probability distribution of energy possessed by atoms at different temperatures is calculated. Atoms in silicon have binding energy due to the covalent bond interaction with their surrounding atoms, and with the irradiation of the femtosecond laser, the atoms are in a high temperature state and have a higher probability of being in a high-energy state. When the atomic energy is higher than the lattice binding energy, the atoms break free from the covalent bonds to form vacancy defects (shown in the inset in Fig. 3a), and then the material cools at an extremely rapid rate, leading to defects freezing. Multiple pulse interaction between femtosecond lasers and the material results in a rapid heating and cooling cycle, where defects are constantly created and frozen, eventually generating a large number of defects inside the material. The defects produced by femtosecond laser irradiation form a bound state near the surface of silicon, and deep defect energy levels in silicon are formed in band gap. These independent deep energy levels eventually lead to broadband absorption in the infrared region, as shown in Fig. 3b, and the schematic diagrams of the 2D and 3D models of the mechanism are shown in Fig. 3c~d.

In order to verify the existence of deep energy levels on the surface of the processed silicon, a high energy femtosecond laser was used to perform a single saturation processing on the surface of the silicon. The processed silicon was cleaned in 10wt% hydrofluoric (HF) solution for 30min to remove the $SiO_2$ generated due to the thermal effect, and then a deep energy level transient spectrometer (DLTS) was used to test the deep energy levels on the surface of the processed area. The corresponding results are shown in Fig. 3f~h. Fig. 3f shows a scanning electron microscope (SEM) image of the sample after cleaning with HF, and the inset indicates that the $SiO_2$ has been removed. Fig. 3g shows the test result of DLTS, where each fitted line represents a deep energy level, and the inset shows the sample preparation structure for DLTS testing. The detailed DLTS sample preparation process and technology are shown in Fig. S6. DLTS data show that the sample after femtosecond laser treatment has 12 deep energy levels, and the band position relationship diagram is shown in Fig. 3h. These deep energy levels are evenly distributed in the region of more than half of the band gap, and each deep energy level can induce the band transition, resulting in broadband optical absorption below its cutoff wavelength (Fig. 3h inset). Band transitions induced by multiple defect levels can achieve a broadband optical absorption covering 3~10μm. Because the optical absorption above 10μm wavelength depends on a shallower energy level (<0.1eV), which is lower than the detection limit of DLTS, it cannot be measured in the experiment.

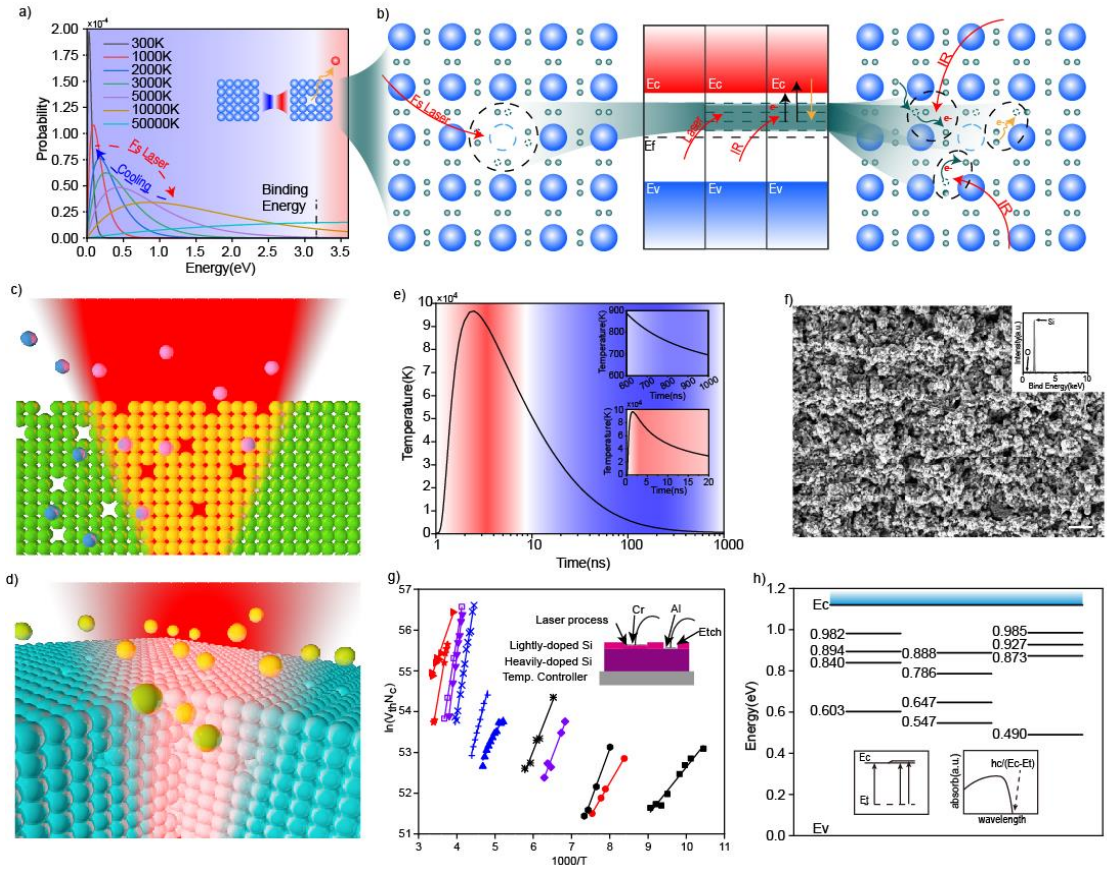

**Fig3 a)** The femtosecond laser heats up the material, and the energy of some atoms exceeds the binding energy to generate defects. **b)** The defect forms a deep energy level to absorb infrared light. **c)~d)** Schematic diagram of laser thermal excitation of defects. **e)** Temperature changes at 200nm below the surface in the laser-affected region. **f)** The SEM image of the laser-affected region and the EDS data (inset diagram). **g)** The result of DLTS and its testing sample structure (inset diagram). **h)** Defect level position after laser irradiation and its principle of optical absorption (inset diagram).

### 3.5. Precise measurement of thin layer absorption

In order to characterize the optical absorption effect of the surface defect layer after processing, we proposed a thin-layer absorption test method. The test process is shown in Fig. 4a~d. In order to avoid the influence of silicon doping on the test results, intrinsic silicon (resistivity > 10000 Ω·cm, 500 μm thickness) is used as the test object, and the test diagram is shown in Fig. S4. Firstly, the intrinsic silicon was placed on the surface of standard absorbing blackbody (the reflectance is nearly 0%) and gold plate (the reflectance is nearly 100%) respectively, and the reflectance $R_1$ and $R_2$ were measured. Then the processed sample was placed on the surface of standard absorbing blackbody and gold plate, and the reflectance $R_4$ and $R_5$ were measured. Based on the reflection physical model we established, the final absorption rate of the thin layer satisfies the following approximate relationship (see detailed derivation process in S7):

$$P = \frac{R_1}{2 - R_1} \quad (E3)$$

$$Q = \frac{R_4 - P}{1 + R_4 P - 2P} \quad (E4)$$

$$X = \frac{1 - R_5 - R_3}{1 - P - \frac{R_3}{2} - Q} \quad (E5)$$

The above steps were used to test the following three samples respectively: the same sample as shown in Fig. 3f, the sample etched with 28wt% KOH solution (60°C, 300s) to remove part of the processing area from the sample in Fig. 3f, and the sample treated with laser at an energy density of 0.134J/cm² (the same processing parameters as Fig. 1i sample) after KOH etching. The test result are shown in Fig. 4e. The results show that the thin layer of the sample surface after the intense laser pulse treatment can absorb more than 70% of the incident light in the wavelength range of 3 ~ 14 μm. At 8 ~ 10 μm, the absorption can reach more than 90%, which is because the interaction between the high energy pulse and material generated $SiO_2$, and $SiO_2$ has a significant absorption peak at 8~10μm. The thin layer absorption after KOH etching is significantly reduced to less than 50%. The reason why it is not reduced to 0 is that the laser-affected region is relatively thick and the defect concentration decreases with depth, so the defects caused by the laser thermal effect cannot be completely eliminated. Due to the low energy density, the irradiated samples (the third sample) cannot generate numerous thermal defects, so the defect absorption cannot be improved on this basis. The change in the range of 8 ~ 10 μm before and after irradiation is due to the $SiO_2$ generated by irradiation.

In addition, the result show that because the defect layer is not thick enough or defect concentration is low, the infrared light cannot be absorbed completely and cause some transmission. So the N-doped silicon used in this work is necessary, because the high level of doping can further absorb the light passing through the laser-processed defect layer, to achieve higher emissivity. In short, the high emissivity is achieved as a result of the combined absorption by the laser-induced defect layer and the doped defects.

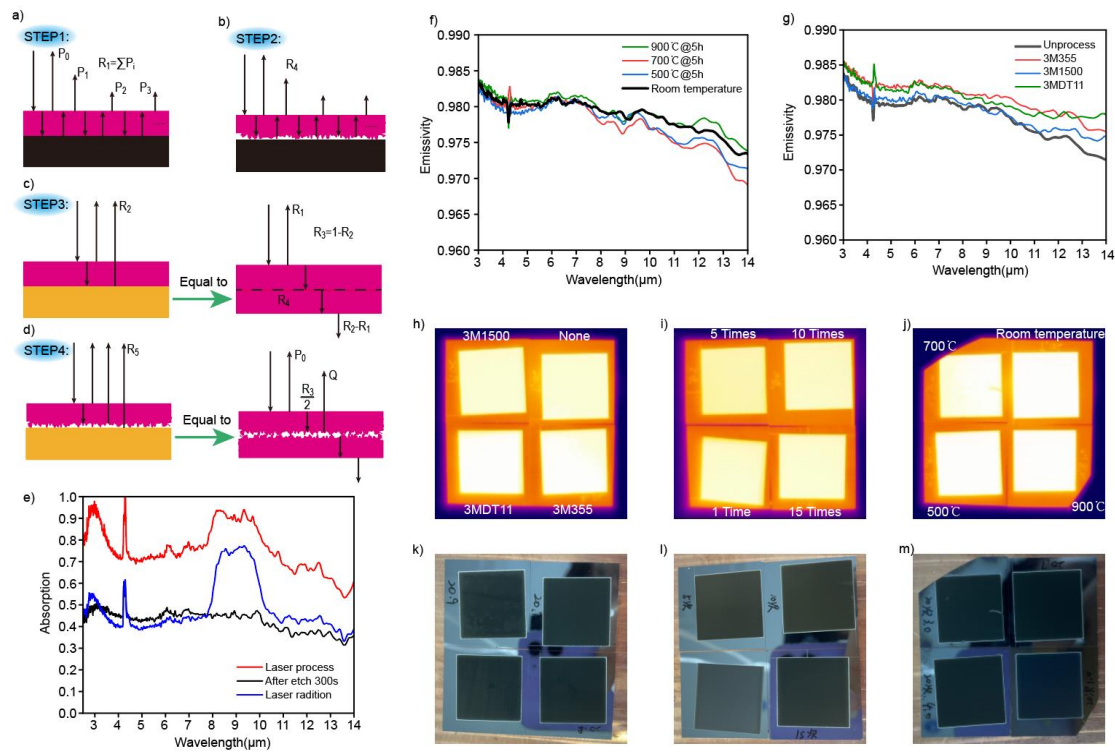

Fig4 a ~ d) The steps of thin layer absorption test. e) Result of thin layer absorption test. f) Result of high temperature stability test. g) Result of stripping test, mechanical stability. h ~ j) Uniformity test. Infrared thermal images of stripping test samples, different process times samples, and high temperature test samples, and its physical photos (k ~ m).

### 3.6. High stability blackbody preparation

A sample with high emissivity was processes using the above method and principle, and its thermal stability and mechanical stability were tested respectively. The test results are shown in Fig. 4f~g. Fig. 4f shows the change of sample emissivity after continuously heating at 500 °C, 700 °C and 900 °C for 5 h, and the sample emissivity is almost unchanged within the wavelength range of 3~14μm, indicating that the sample processed by this method has great thermal stability. In addition, 3M355, 3M1500 and

3MDT11 tapes (produced by 3M company) with different tackiness and materials were respectively used to perform the stripping experiment on the samples (see the physical photo in S8), and the results in Fig. 4g shows that all the samples remain a high emissivity without obvious decrease, indicating that the samples have excellent adhesion and mechanical stability.

In order to verify the emissivity uniformity of the processed and treated samples, we used an infrared thermal imager to perform thermal imaging on all the samples in Fig. 2h and Fig. 4f~g. The results are shown in Fig. 4h~j, and the physical comparison of the corresponding samples are shown in Fig. 4k~m. The results show that the processed samples have good surface uniformity. Even after high temperature or mechanical stripping treatments, although there is a certain non-uniformity in the visible light range, the surface still exhibits excellent uniformity and emissivity in the infrared band. Its uniformity is weakly dependent on the processing parameters, and even if the processing times change obviously, the emissivity uniformity will not be affected. The high stability of processed samples in harsh environments, along with the non-sensitivity and convergence to processing parameters, makes it highly promising for a wide range of application.

## 4. Conclusion

In this work, we proposed to controlling far-field reflectance by adjusting the scattering characteristics of the nanoparticles, and verify that the far-field reflectance can be flexibly controlled by regulating the size of particles. By adjusting the appropriate size, the reflectance can be reduced to much lower than the Fresnel reflectance of the interface, providing a new degree of freedom for the adjustment of interface reflectance. Since the scattering behavior of particles is less affected by the refractive index of materials, the method for adjusting reflectance has a certain material universality, and the effect of adjusting particle size on far-field reflectance in both silicon and silica systems has the same trend. In this work, the physical process is realized by ultrafast laser processing, the experimental and simulation results have an excellent correspondence and expectation.

In addition, this work also proposed a way to further reduce the far-field reflectance by reducing the backscattering integral angle of the scatterer through the microstructure, and proposed a method to accurately control the energy density of low-energy laser by defocusing and a method to achieve the stable convergence of structure after multiple processing, so as to generate a uniform scatterer structure on the surface of the microstructure cone tip. Finally we achieved a reflectance of less than 2% in the infrared wavelength range (3~14 μm), Corresponding to an emissivity exceeding 98%. The fabricated structure has been proved to have great thermal and mechanical stability, as well as high uniformity. The low-reflectance surface processed by the principle and method proposed in this paper has a high emissivity in the entire infrared band of 3~14 μm, and can be rapidly fabricated with a large area by extension. The processed samples can serve as standard radiation sources for high-precision infrared cameras which need space thermal radiation calibration in extreme environments, and they hold promising application prospects in the field of infrared radiometric calibration.

## Acknowledgements

This work was supported by Beijing Natural Science Foundation(L241023), was supported in part by the National Natural Science Foundation of China (NSFC) under grant numbers 62425506.

## Author contributions

H.-S.Zhou, J.-H.Zhang, B.-F.Bai and H.-B.Sun conceived the experiments. H.-S.Zhou, J.-H.Zhang and X.-R.Mei carried out the experiments. H.-S.Zhou and T.Tian performed the numerical simulations. J.S, G.-R.G, X.-P.Hao developed the reflectance measuring system and carried out the measure experiments. H.-S.Zhou, J.-H.Zhang , X.-R.Mei J.-L.Chen and T.Tian developed and improved the fabrication system. H.-S.Zhou, J.-H.Zhang and J.-H.Zhao conceived and carried out the DLTS experiment. B.-F.Bai and H.-B.Sun supervised the whole project. H.-S.Zhou and J.-H.Zhang wrote the initial draught, H.-S.Zhou, J.-H.Zhang and T.T draw all the figures. B.-F.Bai and H.-B.Sun revise the draught, and all authors contributed to the final paper.

**Competing interests**

The authors declare no competing interests.

**Referenced**

1. Sikakwe G. U. Mineral exploration employing drones, contemporary geological satellite remote sensing and geographical information system (GIS) procedures: A review. *Remote Sens Appl-Soc Environ* **31**, 21 (2023).
2. Liu Y., Zuo X. Y., Tian J. F.*, et al.* Research on Generic Optical Remote Sensing Products: A Review of Scientific Exploration, Technology Research, and Engineering Application. *IEEE J Sel Top Appl Earth Observ Remote Sens* **14**, 3937-3953 (2021).
3. Mustard J. F. From planets to crops and back: Remote sensing makes sense. *J Geophys Res-Planets* **122**, 794-797 (2017).
4. Ding J. S., Mi X. L., Chen W.*, et al.* Forecasting of Tropospheric Delay Using AI Foundation Models in Support of Microwave Remote Sensing. *IEEE Trans Geosci Remote Sensing* **62**, 19 (2024).
5. Chen Y., Tao F. L. Potential of remote sensing data-crop model assimilation and seasonal weather forecasts for early-season crop yield forecasting over a large area. *Field Crop Res* **276**, 11 (2022).
6. Ravindra V., Nag S., Li A. Ensemble-Guided Tropical Cyclone Track Forecasting for Optimal Satellite Remote Sensing. *IEEE Trans Geosci Remote Sensing* **59**, 3607-3622 (2021).
7. Michaelchuck E. C., Jr., Lang R. H., Coburn W. O.*, et al.* Non-Plane Wave Incidence on a Body of Revolution for Remote Sensing Applications. In: *Conference on Radar Sensor Technology XXVIII*). Spie-Int Soc Optical Engineering (2024).
8. Lenton T. M., Abrams J. F., Bartsch A.*, et al.* Remotely sensing potential climate change tipping points across scales. *Nat Commun* **15**, 15 (2024).
9. Hutchinson C. F. USES OF SATELLITE DATA FOR FAMINE EARLY WARNING IN SUB-SAHARAN AFRICA. *Int J Remote Sens* **12**, 1405-1421 (1991).
10. Liang S. L., Wang J. D. Atmospheric correction of optical imagery. In:

Advanced Remote Sensing: Terrestrial Information Extraction and Applications, 2nd Edition) (2020).
11. Clough B., Xi-Cheng Z. Toward remote sensing with broadband terahertz waves. *Front Optoelectron (China)* **7**, 199-219 (2014).
12. Franceschetti G., Riccio D. Microwave remote sensing. *URSI 2004 International Symposium on Electromagnetic Theory*, 661-663 vol.662 (2004).
13. Singh R. N. REMOTE-SENSING OF EARTHS RESOURCES USING BROAD-BAND ELECTROMAGNETIC RESPONSE. *Adv Space Res* **13**, 55-64 (1993).
14. Modest M. F. FUNDAMENTALS OF THERMAL RADIATION. In: *Radiative Heat Transfer, 3rd Edition*) (2013).
15. Meseguer J., Pérez-Grande I., Sanz-Andrés A. Thermal radiation heat transfer. In: *Spacecraft Thermal Control*). Woodhead Publ Ltd (2012).
16. Petela R. Exergy of undiluted thermal radiation. *Solar Energy* **74**, 469-488 (2003).
17. Tsakiris G. D. ENERGY REDISTRIBUTION IN CAVITIES BY THERMAL-RADIATION. *Physics of Fluids B-Plasma Physics* **4**, 992-1005 (1992).
18. Sievers A. J. THERMAL-RADIATION FROM METAL-SURFACES. *Journal of the Optical Society of America* **68**, 1505-1516 (1978).
19. Kuenzer C., Zhang J. Z., Dech S. Thermal Infrared Remote Sensing: Principles and Theoretical Background. In: *Remotely Sensed Data Characterization, Classification, and Accuracies*) (2016).
20. Dugan J. P., Anderson S. P., Piotrowski C. C.*, et al.* Airborne Infrared Remote Sensing of Riverine Currents. *IEEE Trans Geosci Remote Sensing* **52**, 3895-3907 (2014).
21. Ng E. Y. K., Acharya R. U. Remote-Sensing Infrared Thermography Reviewing the Applications of Indoor Infrared Fever-Screening Systems. *IEEE Eng Med Biol Mag* **28**, 76-83 (2009).
22. Yamaka E. Infrared remote sensing. *J Jpn Soc Precis Eng (Japan)* **56**, 1984-1989 (1990).
23. Svantner M., Lang V., Skála J.*, et al.* Statistical Study on Human Temperature Measurement by Infrared Thermography. *Sensors* **22**, 19

(2022).

24. Leonidas E., Ayvar-Soberanis S., Laalej H., *et al.* A Comparative Review of Thermocouple and Infrared Radiation Temperature Measurement Methods during the Machining of Metals. *Sensors* **22**, 23 (2022).

25. Maier C. Infrared temperature measurement of polymers. *Polym Eng Sci* **36**, 1502-1512 (1996).

26. Wen Z. D., Zhang Z., Zhang K. P., *et al.* Large-Scale Wideband Light-Trapping Black Silicon Textured by Laser Inducing Assisted with Laser Cleaning in Ambient Air. *Nanomaterials* **12**, 12 (2022).

27. Liu X. L., Radfar B., Chen K. X., *et al.* Perspectives on Black Silicon in Semiconductor Manufacturing: Experimental Comparison of Plasma Etching, MACE, and Fs-Laser Etching. *IEEE Trans Semicond Manuf* **35**, 504-510 (2022).

28. Chen T., Wang W. J., Pan A. F., *et al.* Characterization of anti-reflection structures fabricated via laser micro/ nano-processing. *Opt Mater* **131**, 7 (2022).

29. Fan Z., Cui D. F., Zhang Z. X., *et al.* Recent Progress of Black Silicon: From Fabrications to Applications. *Nanomaterials* **11**, 26 (2021).

30. Zhao J. H., Li X. B., Chen Q. D., *et al.* Ultrafast laser-induced black silicon, from micro-nanostructuring, infrared absorption mechanism, to high performance detecting devices. *Mater Today Nano* **11**, 20 (2020).

31. Zhang Z. X., Martinsen T., Liu G. H., *et al.* Ultralow Broadband Reflectivity in Black Silicon via Synergy between Hierarchical Texture and Specific-Size Au Nanoparticles. *Adv Opt Mater* **8**, 9 (2020).

32. Savin H., Repo P., von Gastrow G., *et al.* Black silicon solar cells with interdigitated back-contacts achieve 22.1% efficiency. *Nat Nanotechnol* **10**, 624-+ (2015).

33. Ge Z. Y., Xu L., Cao Y. Q., *et al.* Substantial Improvement of Short Wavelength Response in n-SiNW/PEDOT:PSS Solar Cell. *Nanoscale Res Lett* **10**, 8 (2015).

34. Saab D. A., Basset P., Pierotti M. J., *et al.* Static and Dynamic Aspects of Black Silicon Formation. *Phys Rev Lett* **113**, 5 (2014).

35. Naik G. V., Shalaev V. M., Boltasseva A. Alternative Plasmonic

Materials: Beyond Gold and Silver. *Adv Mater* **25**, 3264-3294 (2013).

36. Spinelli P., Verschuuren M. A., Polman A. Broadband omnidirectional antireflection coating based on subwavelength surface Mie resonators. *Nat Commun* **3**, 5 (2012).
37. Nguyen K. N., Abi-Saab D., Basset P.*, et al.* Study of black silicon obtained by cryogenic plasma etching: approach to achieve the hot spot of a thermoelectric energy harvester. *Microsyst Technol* **18**, 1807-1814 (2012).
38. Leem J. W., Song Y. M., Yu J. S. Hydrophobic and antireflective characteristics of thermally oxidized periodic Si surface nanostructures. *Appl Phys B-Lasers Opt* **107**, 409-414 (2012).
39. Park B. D., Leem J. W., Yu J. S. Bioinspired Si subwavelength gratings by closely-packed silica nanospheres as etch masks for efficient antireflective surface. *Appl Phys B-Lasers Opt* **105**, 335-342 (2011).
40. Huang G., Yengannagari A. R., Matsumori K.*, et al.* Radiative cooling and indoor light management enabled by a transparent and self-cleaning polymer-based metamaterial. *Nat Commun* **15**, 12 (2024).
41. Reicks A., Tsubaki A., Anderson M.*, et al.* Near-unity broadband omnidirectional emissivity via femtosecond laser surface processing. *Commun Mater* **2**, 11 (2021).
42. Du M. D., Sun Q. Q., Jiao W.*, et al.* Fabrication of Antireflection Micro/Nanostructures on the Surface of Aluminum Alloy by Femtosecond Laser. *Micromachines* **12**, 11 (2021).
43. Wenzheng Z., Daozhi S., Guisheng Z.*, et al.* Super black iron nanostructures with broadband ultralow reflectance for efficient photothermal conversion. *Appl Surf Sci* **521**, 146388 (146386 pp.)-146388 (146386 pp.) (2020).
44. Lou R., Zhang G. D., Li G. Y.*, et al.* Design and Fabrication of Dual-Scale Broadband Antireflective Structures on Metal Surfaces by Using Nanosecond and Femtosecond Lasers. *Micromachines* **11**, 11 (2020).
45. Liu H. L., Hu J., Jiang L.*, et al.* Ultrabroad antireflection urchin-like array through synergy of inverse fabrications by femtosecond laser-tuned chemical process. *Appl Surf Sci* **528**, 9 (2020).
46. Jalil S. A., Lai B., ElKabbash M.*, et al.* Spectral absorption control of

femtosecond laser-treated metals and application in solar-thermal devices. *Light, science & applications* **9**, 14 (2020).
47. Su Y. L., Hu H., Feng H.*, et al.* A novel generation scheme of ultra-short pulse trains with multiple wavelengths. *Opt Commun* **389**, 176-180 (2017).
48. Fan P. X., Bai B. F., Zhong M. L.*, et al.* General Strategy toward Dual-Scale-Controlled Metallic Micro-Nano Hybrid Structures with Ultralow Reflectance. *ACS Nano* **11**, 7401-7408 (2017).
49. Zhou L., Tan Y. L., Wang J. Y.*, et al.* 3D self-assembly of aluminium nanoparticles for plasmon-enhanced solar desalination. *Nat Photonics* **10**, 393-+ (2016).
50. Zhou L., Tan Y. L., Ji D. X.*, et al.* Self-assembly of highly efficient, broadband plasmonic absorbers for solar steam generation. *Sci Adv* **2**, 8 (2016).
51. Fan P. X., Bai B. F., Long J. Y.*, et al.* Broadband High-Performance Infrared Antireflection Nanowires Facilely Grown on Ultrafast Laser Structured Cu Surface. *Nano Letters* **15**, 5988-5994 (2015).

# Supplementary Information

# Contents:

# S1.The Fresnel reflection

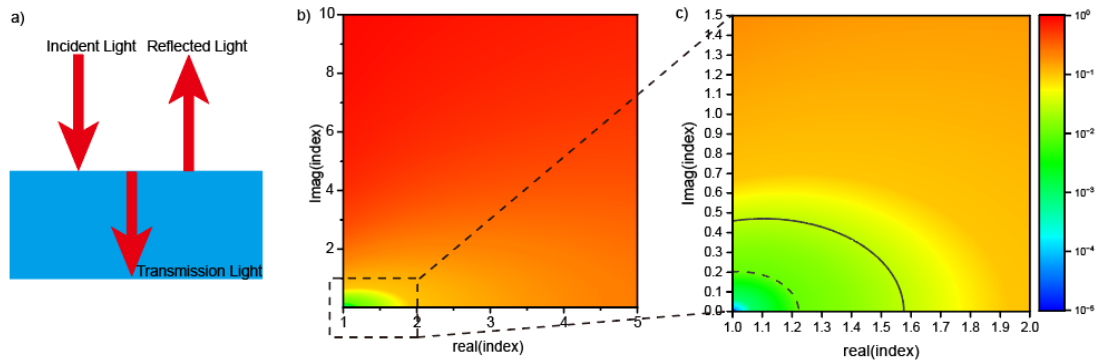

a) The reflection behaviors' schematic at the interface of two materials. b) The surface reflectance of materials in air given by Fresnel's law. The transverse axis is the real part of refractive index, and the longitudinal axis is the imaginary part of refractive index. c) The enlarged of b). The solid line is the 5% reflectance, and the dashed line is the 1% reflectance.

Light reflected at the interface as shown in a), and the reflection characteristics of vertically incident unpolarized light are given by the following Fresnel formula:

$$R = \frac{(n_r - 1)^2 + n_i^2}{(n_r + 1)^2 + n_i^2}$$

Where the $n_r$ is the real part of refractive index, the $n_i$ is the imaginary part of refractive index. b) and c) give the reflectance calculated by this formula. If reflectance reaches 1%, it needs the refractive index lower than about 1.2+0.2i, and if reaches 5%, it needs that lower than about 1.5+0.4i. For commonly used materials, it is stringent.

## S2.Ultrafast laser surface processing system

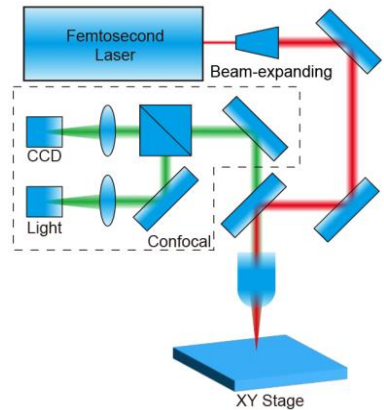

The figure shows the femtosecond laser processing system. It consists of a beam expanding, which is used for enlarging the beam diameter; an objective, used for focusing the laser to materials; a X-Y-Z displacement stage, X-Y used for moving the samples to process in specific path, Z used for focusing or defocusing; the confocal system, used for online observation.

# S3.The principle of defocusing energy control for Gaussian beam

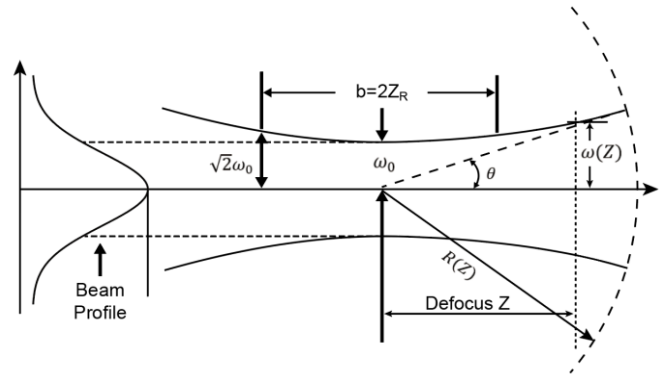

Figure shows the characteristics diagram of Gaussian beam, its properties are given by the following equation set:

$$N_A = n\sin\theta$$

$$\omega_0 = \frac{\lambda}{\pi\theta}$$

$$\omega(z) = \omega_0 \sqrt{1 + \left(\frac{\lambda z}{\pi\omega_0^2}\right)^2}$$

$$I(z,r) = \frac{2P}{\pi\omega(z)^2} \exp\left(-\frac{2r^2}{\omega(z)^2}\right)$$

Where $N_A$ is the numerical aperture of the objective lens, $\omega_0$ is the beam waist radius at the focus, $\omega(z)$ is the beam radius at position z, $I(z,r)$ is the intensity distribution, $\lambda$ is the wavelength of laser. For a defocus position z, the beam spot area is given by the following formula:

$$S = \pi\omega(z)^2 = \pi\omega_0^2 + \frac{(\lambda z)^2}{\pi\omega_0^2}$$

So, if $\lambda z \gg \omega_0$, and the pulse energy is E, the pulse energy density is given by the following formula:

$$\rho_S = \frac{E}{S} = \frac{\pi\omega_0^2 E}{(\lambda z)^2}$$

If the laser power is P, the pulse repetition frequency is F, then the $\rho_S$ is given by:

$$\rho_S = \frac{E}{S} = \frac{\pi\omega_0^2 P}{(\lambda z)^2 F}$$

For this work, P=4 J/s, F=$10^5$ /s, $N_A = 0.14$, $\lambda = 1.03\ \mu m$, so the

defocus distance z determines the beam waist radius and pulse energy density:

$$\omega_0 = 2.33 \ \mu m$$

$$\rho_S = \frac{3.14 * (2.33 \ \mu m)^2 * 4 \ J/s}{1.06 \ \mu m^2 * 10^5 \ /s * z^2} = \frac{6.433 * 10^{-4} \ J}{z^2}$$

The following table gives the pulse energy density calculated by above formulas which used in this work:

| Z(cm) | $\rho_S (J/cm^2)$ |
|---|---|
| 0.015 (150 μm) | 2.840 |
| 0.070 (700 μm) | 0.131 |
| 0.080 (800 μm) | 0.101 |
| 0.090 (900 μm) | 0.079 |
| 0.100 (1000 μm) | 0.064 |

## S4.The method of reflectance testing

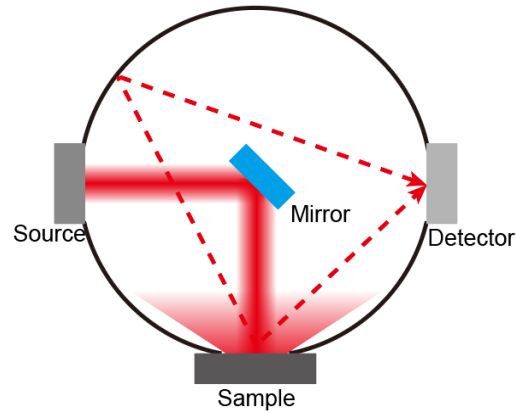

　　　Figure shows the schematic diagram of reflectance measurement. An integrating sphere with a gold-coated inner wall　(diffuse the infrared light, without absorption or transmission) was used for hemispheric reflectance measurement. A gold-plated mirror was used for reflecting the light from infrared source to the sample. A detector was used for detecting infrared light. A Fourier infrared spectroscopy (FTIR) was used for analyzing and calculating the reflectance.

## S5. The XPS/EDS data of the processed samples

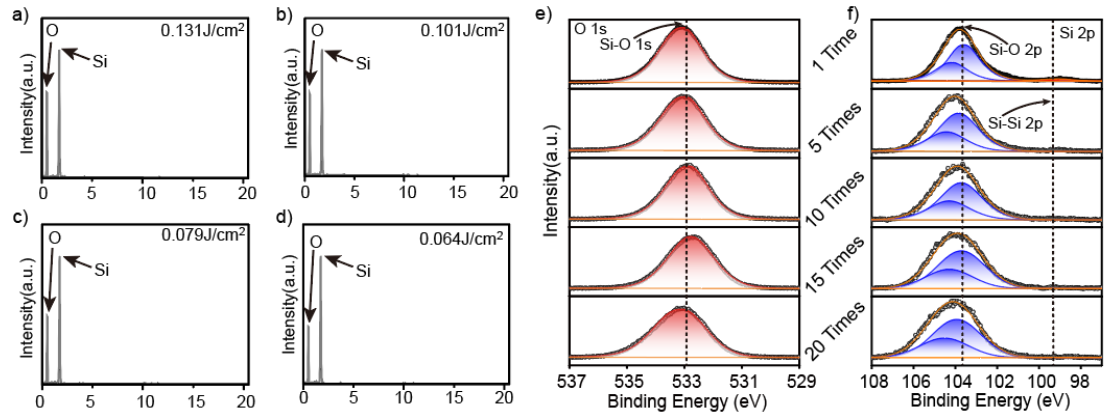

a ~ d) The EDS data corresponding to the samples of Fig. 1k in main text. The percentage of Si and O shows that the SiO2 and pure element Si exist on the surface of processed region. e) ~ f) The XPS data corresponding to the sample of Fig. 2g in main text. It shows that the chemical composition of the surface microstructure formed by multiple processing is the same as that of the surface nanoparticles formed by irradiating with low energy density.

## S6. Sample preparation process of DLTS

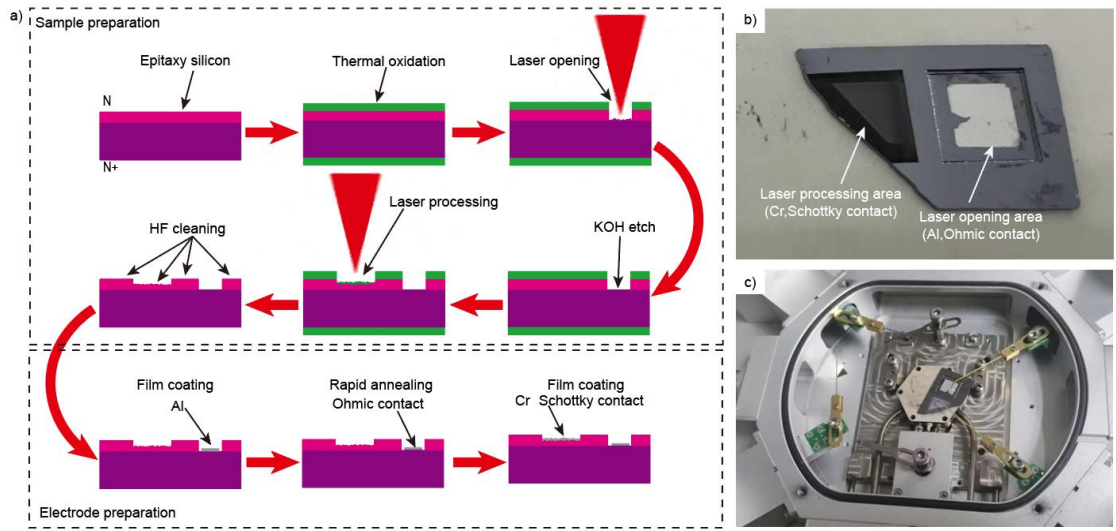

a) The sample preparation steps for DLTS test. There are two main steps, the sample preparation and the electrode preparation. b) Physical photo of DLTS sample. c) DLTS testing environment.

DLTS detects the defect energy levels by testing the gradient of junction capacitance variation with voltage changes, so a p-n junction or metal-semiconductor junction is necessary. It is difficult to generate a p-n junction on the laser-affected region, but the metal-semiconductor junction is easier. In addition, the semiconductor electrode needs a ohmic contact to insure smaller gradient of voltage and capacitance to avoid influencing the test.

Fig a) shows the preparation steps for DLTS sample. At first, a heavily doped silicon wafer with a lightly doped epitaxial (110 μm) is used as the laser processing materials. Then, the silicon sample is thermally oxidized (1100 ℃, 5h) to generate a silica layer, as the barrier layer for the following KOH etching that prevents the silicon surface from being corroded by KOH. Subsequently, we used a high-power femtosecond laser scanning on the surface region of epitaxy layer, to remove the $SiO_2$ and some epitaxy Si in that region, as the window. And the KOH etching (28w%, 60 ℃, 1h) was performed to remove the defects generated by laser, further deepening and chemically polishing the window area. Then high-power femtosecond laser scanned the materials to generate the defect

region (the target region outside the window region), and HF was used for cleaning the silica generated by thermal oxidation and laser. So far, the sample preparation finished.

A film of Al about 1μm was coated on the opening window region (heavy doping silicon) and a rapid annealing was used for generating the Al-Si eutectic compound to form the ohmic contact, in order to avoid the electrode contact potential difference which affects test results. After this, Cr film was coated on the target testing region to form metal-semiconductor contact (Schottky junction). Fig b) shows the sample photograph prepared by the above process. Fig c) shows the DLTS instrument used in this test.

# S7. The deduce of precise measurement of thin layer absorption

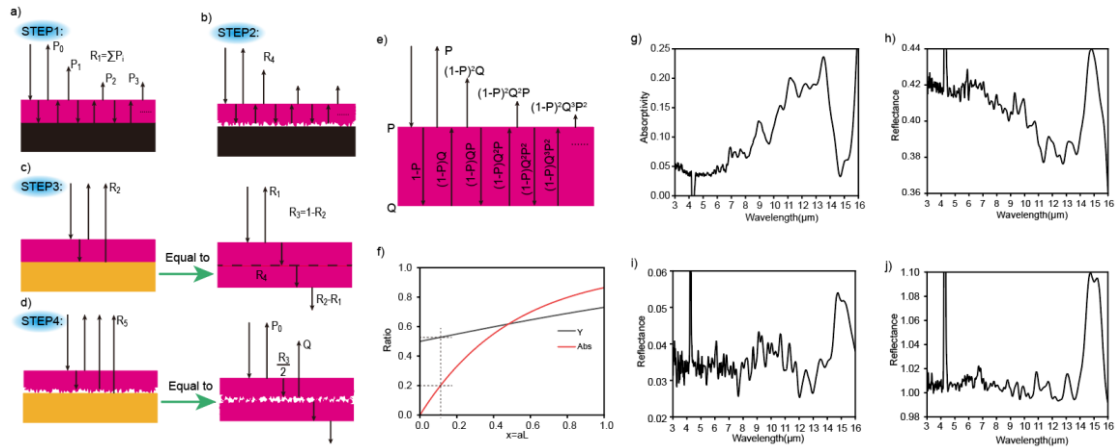

**a~d) The process of testing. e) The theory model of multiple oscillation reflection. f) The functions curve of Lambert-Beer correction coefficient. g) The absorptivity of polished Si. h) Reflectance of polished Si ($R_1$). i) the reflectance of standard blackbody. j) the reflectance of standard Au plate.**

Figure a) ~ d) (Fig. 4a~d) shows the measurement process of thin layer absorption. We consider an air-film-air (three layers) materials system. Both the upper surface of the upper layer (air) and the lower surface of the lower layer (air) connect to the perfectly matched layer (PML), so ignoring the reflection in upper and lower layers. We only consider the reflection in the film layer, including the interfaces between upper and middle layers, lower and middle layers, as shown in Fig e). If the reflectance on the interface between upper layer and middle layer is P, on the surface between lower layer and middle layer is Q, when a beam of light incident, first reflection intensity is P, so the intensity of (1-P) enter the middle layer, and when it reaches the next interface, the Q of (1-P) was reflection intensity, i.e. (1-P)Q. When it secondly reaches the first interface, (1-P) of (1-P) Q was transmission intensity and P of (1-P) Q was reflection intensity, i.e. $(1-P)^2Q$ and (1-P) QP, so Infinite times oscillation occur in the middle layer, just like a F-P cavity. Each time the transmission intensity from upper interface is as follow:

$$P_0 = P$$

$$P_1 = (1 - P)^2 Q$$
$$P_2 = (1 - P)^2 Q^2 P$$
$$P_2 = (1 - P)^2 Q^3 P^2$$
$$\ldots \ldots$$
$$P_2 = (1 - P)^2 Q^3 P^2$$
$$P_n = (1 - P)^2 Q Q^n P^n$$

$$R = P + \sum_{n=1}^{\infty} (1 - P)^2 Q Q^n P^n = P + \frac{2(1 - P)Q}{1 - PQ} \qquad S7.1$$

If the upper and the lower layers are the same material, then P=Q, so:

$$R = \frac{2P}{1 + P}$$

The test of Fig a) is a polished silicon onside the standard blackbody, conforming the condition of above formula, so we can measure the Si-Air surface reflectance as follows:

$$P = \frac{R_1}{2 - R_1} \qquad S7.2$$

Next, we consider a rough surface processed by laser, which was placed upside down on the standard blackbody, shown as Fig b), and now, $P \neq Q$. Because we measured the $R_1$, we can calculate $P$, so we can obtain equivalent reflectance $Q$ of the rough surface as follows:

$$Q = \frac{R_4 - P}{1 + R_4 P - 2P} \qquad S7.3$$

Where the $R_4$ was measured in step 2, shown in Fig b).

Because we want to measure the absorption of defects, the next step we put the polished Si onside the standard Au plate, shown as Fig c), which can be equal to a double thickness layer, and if the measured value in step 3 is $R_2$, the polished Si absorption (double thickness) is as follows:

$$R_3 = 1 - R_2$$

Where the $R_3$ is the absorptivity of double thickness polished Si.

Now we get all the parameters of the polished Si, then we consider replace the polished Si with the processed Si, which was still placed upside down on the Au plate as shown in Fig d). We consider the equal figure, the beam of light firstly was reflected (P), and about $R_3/2$ was absorbed. The remaining light reached the rough surface and Q of them was reflected, some of the others was absorbed (absorptivity is X), and the final remaining part passed through the symmetrical Si and another $R_3/2$ was absorbed. Finally we measured the value $R_5$. According to the path of light, the following equation is established:

$$\left(1 - P - \frac{R_3}{2} - Q\right)X + R_3 = 1 - R_5$$

Where the first item on the left is the part absorbed by defect layer, the second item is the part absorbed by background Si. The right item was all the absorption, in which the $R_5$ was the measured value from step 4. Thus, it can be deduced:

$$X = \frac{1 - R_5 - R_3}{1 - P - \frac{R_3}{2} - Q} \qquad S7.4$$

So far, the defect absorption was measured by the above four steps.

In the deduce of S7.4, we assumed that the optical absorption of half a layer is equal to half of the optical absorption of the total layer, which needs to confirm. We consider the absorption of a layer with 2L thickness, whose transmittance conforms to the Lambert-Beer law as follows:

$$T = e^{-a2L}$$

So, the absorptivity of it meets:

$$Abs = I = 1 - T = 1 - e^{-a2L}$$

And if the thickness of layer is L, the absorptivity of it is as follows:

$$I' = 1 - e^{-aL}$$

So, we define the Lambert-Beer correction coefficient Y as follows:

$$Y = \frac{I'}{I} = \frac{1 - e^{-aL}}{1 - e^{-a2L}} = \frac{1}{1 + e^{-aL}}$$

If we define $x = aL$, so

$$\begin{cases} Y = \dfrac{1}{1 + e^{-x}} \\ Abs = 1 - e^{-2x} \end{cases}$$

Fig f) shows these function curves when $x$ is in the range from 0 to 1, and Fig g) gives the absorptivity of polished Si, in which the peak absorptivity is about 0.2. We take a point with the ordinate of 0.2 on the curve of Abs, and draw a line parallel to the vertical axis, which intersects

with the curve Y over one point. Then the ordinate of the intersection point is the Lambert-Beer correction coefficient. We can see that it approaches 0.5, meaning it is rational that we assume the absorption of half a layer is approximately equal to half absorption of the total layer.

Fig i) and j) give the reflectance of standard blackbody and standard Au plate which were used in our test. Their measurement accuracy error is approximately 3%.

**In this work, the test method has some other systematic errors and approximation, but we won't discuss it in detail. We only pay attention to the variation trend and the comparative trend. Under the conditions of qualitative analysis, ignoring and approximating these errors can be accepted.**

## S8. The stripping experiments

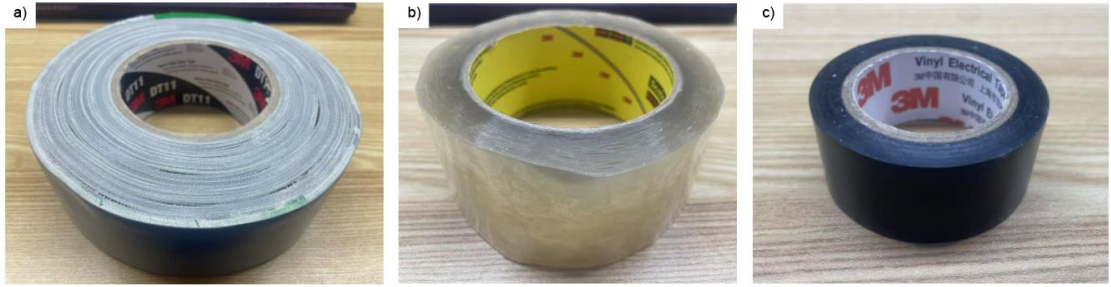

**3MDT11; b) 3M355; c) 3M1500.**

    The adhesive tapes are used for stripping experiments to measure the mechanical stability as shown in above figure. 3MDT11 is a cloth tape with high tackiness, 3M355 is a plastic tape with low tackiness, and the 3M1500 is an insulating tape with the middle tackiness.